\documentclass[runningheads, a4paper]{llncs}
\usepackage{cite}
\usepackage{amsmath,amssymb,amsfonts}
\usepackage{algorithmic}
\usepackage{graphicx}
\usepackage{textcomp}
\usepackage{bmpsize}
\usepackage{xcolor}
\usepackage{lipsum}
\usepackage{multirow}
\usepackage{subcaption}
\usepackage{comment, float, balance, tabularx, pifont}
\usepackage[ruled,vlined,linesnumbered]{algorithm2e}
\usepackage{hyperref}
\usepackage[normalem]{ulem}
\usepackage{tikz}
\usepackage{bbding}

\usetikzlibrary{chains,shapes.multipart}
\usetikzlibrary{shapes,calc}
\usetikzlibrary{automata,positioning}

\SetCommentSty{mycommfont}
\useunder{\uline}{\ul}{}

\newcommand{\cmark}{\ding{51}}
\newcommand{\xmark}{\ding{55}}

\begin{document}

\title{The Dark (and Bright) Side of IoT: Attacks and Countermeasures for Identifying Smart Home Devices and Services}

\titlerunning{The Dark (and Bright) Side of IoT: Attacks and Countermeasures}

\author{Ahmed Mohamed Hussain\inst{1(}\Envelope\inst{)} \and Gabriele Oligeri\inst{2} \and Thiemo Voigt\inst{1}}

\authorrunning{A. M. Hussain et al.}

\institute{
Department of Information Technology, \\Uppsala University, Uppsala, Sweden \and
Division of Information and Computing Technology (ICT), \\College of Science and Engineering (CSE), \\Hamad bin Khalifa University (HBKU)\\
Email: \email{ahmed.hussain.7023@student.uu.se}
}

\maketitle

\begin{abstract}
    We present a new machine learning-based attack that exploits network patterns to detect the presence of smart IoT devices and running services in the WiFi radio spectrum. We perform an extensive measurement campaign of data collection, and we build up a model describing the traffic patterns characterizing three popular IoT smart home devices, i.e., Google Nest Mini, Amazon Echo, and Amazon Echo Dot. We prove that it is possible to detect and identify with overwhelming probability their presence and the services running by the aforementioned devices in a crowded WiFi scenario. This work proves that standard encryption techniques alone are not sufficient to protect the privacy of the end-user, since the network traffic itself exposes the presence of both the device and the associated service. While more work is required to prevent non-trusted third parties to detect and identify the user's devices, we introduce \emph{Eclipse}, a technique to mitigate these types of attacks, which reshapes the traffic making the identification of the devices and the associated services similar to the random classification baseline.
\end{abstract}

\section{Introduction}
Security and privacy of communications are of paramount importance given the growing number of user devices connected to the Internet. Indeed, the number of smart devices is exponentially growing around the world, and the vast majority of today's network traffic is encrypted to guarantee the security and privacy of the communications. Recently, encrypted network traffic classification has emerged as a technique to detect and identify traffic patterns although being anonymized by one or more encryption layers~\cite{li2007machine, shahid2018iot, sivanathan2016low}. The same techniques have been successfully adopted for drones~\cite{SCIANCALEPORE2020107044} and cryptojacking detection~\cite{caprolu2019cryptomining}.
The main idea behind traffic classification is to characterize the network traffic flow according to statistical features independently of the information that can be retrieved from the packet headers, such as network and physical layer addresses. Standard features can be packet size and interarrival times between consecutive packets, while more advanced features take into account statistics such as mean and variance computed over sequences of consecutive packets.
While HTTPS and VPN services protect the end user from adversaries willing to exfiltrate information from the content of the packet, traffic patterns are very difficult to hide without affecting services' performance. Moreover, the wireless scenarios considered in this paper make such attacks even more effective and efficient. Indeed, due to its intrinsic broadcast nature, wireless communications can be easily eavesdropped without the consent of the user, and an adversary can exploit the traffic pattern generated by users to infer either the devices they have at home or the services they are using. This presents major issues for the users' privacy, not just because non-trusted third parties can infer the user behavior, but also detect the presence of specific devices to target in subsequent attacks.

\noindent{\bf Contribution.} We propose a methodology based on machine learning classification to detect and identify the presence of smart devices and their running services in the WiFi radio spectrum. We tested our approach against three different smart devices, i.e., Google Nest, Amazon Echo and Amazon Echo Dot, and three popular services, i.e., Music, YouTube and News streaming. We proved that the aforementioned devices and services are affected by a critical privacy leakage since they allow not-authorized third parties to detect their presence even in the case of standard multi-layer encrypted streams such as WPA and HTTPS. Additionally, we construct and release the data-set\footnote{\url{https://github.com/AMHD/The-Dark-and-Bright-Side-of-IoT-Dataset}} associated with each service and device used in this paper.
This paper proves that the most commonly used smart devices in the market and their related services can be easily detected with overwhelming probability ($>0.99$) from a crowded WiFi link. We provide deep statistical insights about the traffic characterization, we show how standard machine learning techniques can be used to detect the presence of the devices. Finally, we introduce a mitigation technique called "Eclipse", which reshapes the traffic making the identification of the devices and the associated services similar to a random guess.

\noindent{\bf Paper Organization.} The remainder of this paper is organized as follows: Section~\ref{sec:related_work} summarizes recent contributions in the field of encrypted traffic classification, while Section~\ref{sec:advModel} illustrates our assumptions and the adversary model. Our measurement setup and data collection procedures are introduced in Section~\ref{sec:measurement_setup}, while Section~\ref{sec:data_processing_classification} introduces the statistical analysis and reports the performance of our device identification methodology. Section~\ref{sec:device_detection} shows the results of our detection algorithm when considering a crowded WiFi link, and Section~\ref{sec:countermeasure} discusses countermeasures against our attack. Finally, Section~\ref{sec:conclusions} tightens conclusions and draws some future work.

\section{Related Work}
\label{sec:related_work}
Msadek et al.~\cite{msadek2019iot} proposed to fingerprint and map IoT devices to their associated encrypted traffic flow using machine learning. They assumed that the adversary is capable of passively collecting the traffic generated from the IoT devices through monitoring the gateway or a compromised IoT device. 
The data-set used in their analysis~\cite{iotUNSW} constituted of seven different categories. The adversary in their study is remote and capable of obtaining the bidirectional flows generated by the devices. They reported the performance from five different classification algorithms, where \emph{Adaboost}~\cite{jan2018ada} outperformed with 95.5\% accuracy in identifying the type of the device. However, their analysis did not focus on either applications or services these devices are running.

Shahid et al.~\cite{shahid2018iot} presented a machine learning-based approach to identify the type of four different IoT devices connected to the network, by analyzing the bidirectional flows. From the first $N$ packets exchanged between the internet and IoT devices, they extracted two features: packet size and inter-arrival times. Devices were connected to an access point and traffic between the access point and internet is redirected through a Raspberry Pi for capturing packets. Different supervised machine learning algorithms were used for classification, best performance has been achieved using Random Forest with 99.9\% accuracy. Their methodology involves collecting the data from an intermediate communication node (Raspberry Pi) and not directly from the wireless radio spectrum as we do.

Santos et al.~\cite{santos2018efficient} used a data-set that was collected within a smart campus~\cite{sivanathan2017characterizing}. The total number of IoT and non-IoT devices in this data-set is 18. The considered features were maximum packet size in the forward direction, source port, destination port, and average packet size. They were able to identify each device network traffic flow, with 99\% of accuracy by using the Random Forest algorithm. Even in this case study, the authors were able to distinguish devices, but not the application run by the device itself. 

Sivanathan et al.~\cite{sivanathan2017characterizing} considered a data-set collected over a period of three weeks, for more than 20 IoT devices such as cameras, lights, appliances, and health monitors. They extracted features such as data rates and burstiness, activity cycles, and signaling patterns. By exploiting the aforementioned features, they were able to distinguish IoT from non-IoT traffic, furthermore identifying specific IoT devices with about 95\% accuracy. Similarly to previous related work, this contribution focuses on the device but not on the applications run by the device. 

Jackson and Camp~\cite{jackson2018amazon}~investigated the performance of six different supervised machine learning algorithms to infer information from encrypted TCP packets. They were able to identify encrypted information/request\footnote{Such as: "Alexa, turn off the light"} exchanged between the Alexa cloud service and the Echo device by only looking at the request packets exchanged. They adopted three different feature vectors: TCPtrace Features Vector~\cite{lazarevic2003comparative}, Histogram Feature Vector, and Combined Feature Vector. They were able to achieve an accuracy of more than 93\% using the Random Forest algorithm for each of the features vectors. Data collection set-up in this study involves an intermediate communication node (computer) and not directly from the wireless radio spectrum as we propose in this paper.

Trimananda et al.~\cite{trimananda2019pingpong}~developed a tool that fingerprints generated traffic by smart home IoT devices when switched on or off, through analyzing the initial exchanged packets. This solution is effective only at the startup process, while it does not provide any guarantees when the eavesdropped packets comes from an arbitrary point of the stream.

Valdez et al.~\cite{8815667} focused on the identification and classification of IoT devices network traffic, while being encrypted. Leveraging the identify key features of the TLS handshake protocol, they were able to build a model that is capable of identifying 71 IoT devices, with an accuracy over 90\%.

Different methods has been proposed in literature to prevent traffic detection, most notably: traffic and packet \emph{padding}~\cite{chen2010side, sun2002statistical}, traffic \emph{morphing}~\cite{8730787}, and \emph{frequency hopping}. These techniques do mitigate the traffic pattern, but mainly focusing only on one feature (notably packet size). On the other hand, the \emph{Leaky Bucket} algorithm can be used to mitigate the pattern created by the interarrival times by reshaping the traffic at a constant rate.

\section{Adversarial Model}
\label{sec:advModel}
The main reason for performing this attack on the WiFi link, is due to the effectiveness and how easy it is to sniff (collect) the traffic from the WiFi spectrum between the access point and devices without the need to be connected to any of the sending /receiving parties. In this attack, we are particularly interested in services that generates Quality of Service (QoS) traffic, as it generates large number of packets exchanged between the devices and the access point, unlike switching on/off IoT devices.

\begin{figure}[h]
    \centering
    \includegraphics[width=.8\textwidth]{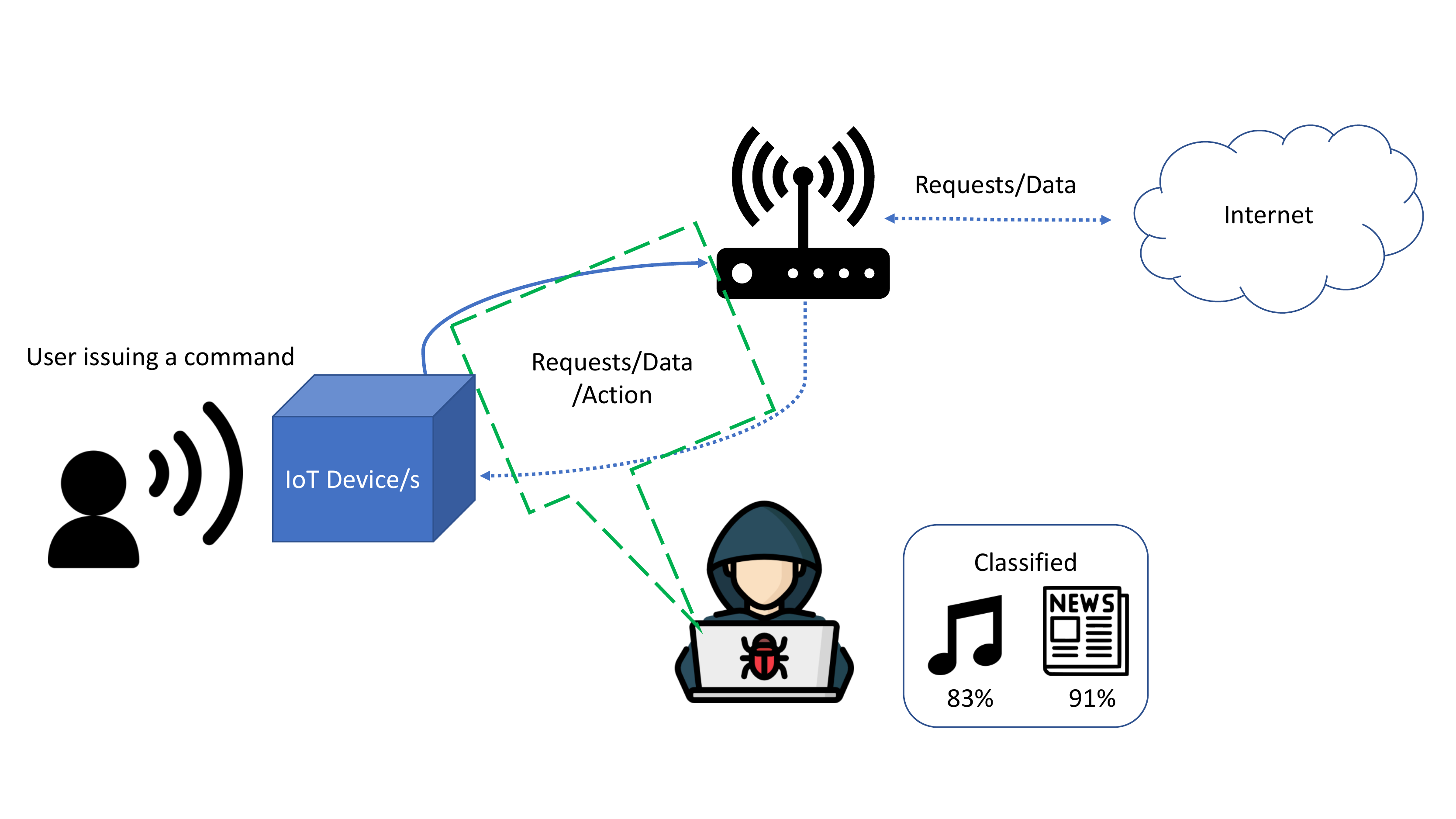}
    \caption{Adversary model considered in this work.}
    \label{fig:attack_setup}
\end{figure}

Our adversary model features two essential characteristics: \emph{stealthy} and \emph{resource constrained}. Indeed, we assume that our adversary is able to eavesdrop and collect 802.11 radio messages (WiFi) from a remote location, even far away from the target user depending on the receiving capabilities of his equipment. Figure~\ref{fig:attack_setup} wraps up our assumption on the adversarial model. We consider a general user case scenario, where the user enjoys different IoT/smart devices connected to the Internet. The aforementioned devices provide different smart services such as news reading and music streaming by resorting to the user's voice commands. The voice command is detected and processed by the smart device, and in turn, sent to the cloud service. Subsequently, the cloud service replies to the smart device with the answer that (in this work) consists of an audio stream. 

We assume the adversary is sitting far away from both the device and the WiFi access point, but still, he can eavesdrop and collect the WiFi radio messages. In particular, our adversary model takes into account the stream of messages from the cloud to the device (ingoing flow). For each eavesdropped message, the adversary records only the packet size and the time at which the packet is received.

We stress that our analysis is powerful even in the presence of a user able to anonymize all the messages, e.g., by sanitizing all the information related to source and destination IP/MAC addresses. Although we recognize that this will require the deployment of ad-hoc protocols able to guarantee the message delivery between the smart device and the WiFi access point, we do assume the worst possible conditions from the adversary perspective, i.e., our adversary will not resort to standard techniques to infer on the service/device such as port scanning or MAC addressing manufacturer mapping.

\vspace{-0.3cm}

\vspace{-0.3cm}

\section{Measurement Set-up and Data collection}
\label{sec:measurement_setup}
In this section, we introduce our measurement set-up by considering both the hardware and the software tools we adopted.

\vspace{0.2cm}  

\noindent{\bf Hardware--Smart Devices.} Our smart home set-up consists of three smart home devices: \emph{Amazon Echo}, \emph{Amazon Echo Dot}, and \emph{Google Nest Mini}. These devices are distributed in different places and connected to a D-Link router (1200AC) via WiFi. 

\vspace{0.2cm}  

\noindent{\bf Adversary--Hardware.} We adopted an Alfa Card (AWUS036NH) for capturing the traffic by switching to \emph{monitor mode}, and therefore, collecting all the over-the-air data traffic.

\vspace{0.2cm}  

\noindent{\bf Adversary--Software.}  We consider a standard Linux distribution (Kali Linux), and the tools \emph{airodump-ng}/\emph{Wireshark} for collecting and logging the WiFi packets. Finally, we adopted \emph{MATLAB 2019b} for data preprocessing and classification.

\vspace{0.2cm}  

\noindent\textbf{Data collection.} We would like to stress that we resort to MAC addresses to select the traffic of the smart devices and generate the ground truth with the machine learning algorithms. As it will be clear in the following sections, the classification will be independent of the aforementioned information, while being rooted only on features like packet size, interarrival times and statistical computations derived from them.

\noindent The data collection has been performed accordingly to the following procedure:

\begin{enumerate}
    \item We adopted \emph{airodump-ng} to identify the victim access point and the devices connected to it. 
    \item Identifying information such as MAC addresses and WiFi channel used by the devices in the network.
    \item Setting the Alfa card channel to the one used by the smart device.
    \item Using either \emph{TCPDump} or \emph{Wireshark} for traffic collection and filtering out the traffic generated from the access point to each of the IoT device, using the MAC address obtained from step 1.
    \item Extract the time and packet size associated with each over-the-air message.
\end{enumerate}

The traffic forwarded by the access point to each of the smart devices is collected over approximately 8 hours of continuous news and music streaming. In our set-up, we ensure that no other devices are connected to the network other than the IoT devices. The total number of packets forwarded by the WiFi access point to each device is approximately 280,000 for each service. In total, the overall data-set size sums up to 1,680,000 packets of streaming news on amazon echo and echo dot and streaming music on echo, echo dot, and google nest mini.

\section{Data Processing, Statistics, and Device Identification}
\label{sec:data_processing_classification}
In this section, we provide the analysis, methodologies and techniques we adopted for \emph{pre-processing} and \emph{classifying} the collected traffic.

\vspace{0.5cm}

\noindent{\bf Pre-processing.} The raw reception times and packet sizes are extracted from the traces. Inter-arrival times are calculated as $T_{x + 1} - T_{x}$, where $T_{x+1}$ is the time of the packet $x+1$ that is received after packet $x$. After this preliminary computation, 8 different features are computed considering different \emph{window sizes} from the inter-arrival times and packet sizes. 

\noindent Different statistical features have been adopted in the literature to uniquely identify patterns from encrypted traffic~\cite{jackson2018amazon, acar2018peek, santos2018efficient}. In this work, we consider 8 statistical features: standard deviation, sum, variance, maximum, minimum, mean, median, skewness, and kurtosis. These features are extracted by using a sliding window technique, where a window of \emph{N} adjacent massages is considered for the computation of the aforementioned statistics.

\vspace{0.5cm}

\noindent{\bf Classification.} Machine learning algorithms facilitates its capabilities to identify and distinguish unique patterns from the used features (interarrival times and packet sizes), which result in identifying services and devices more accurately. The vast majority of our results adopts the Random Forest algorithm for all our tests being the best performer from related works in the literature~\cite{msadek2019iot, acar2018peek, jackson2018amazon, shahid2018iot, santos2018efficient, sivanathan2017characterizing}.
We consider the following configuration for the random forest algorithm: \textbf{k-folds} = 15, \textbf{number of trees/bags} = 30. All classes have been balanced to ensure that the number of samples is equal among all the traces.

\subsection{Interarrival Times Analysis}
\label{sec:intertime}
\begin{figure}[H]
    \centering
    \includegraphics[width=.7\textwidth]{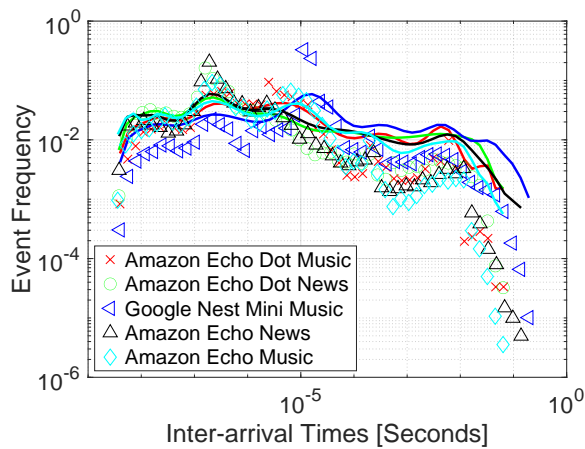}
    \caption{Probability distribution function associated with the interarrival times (seconds).}
    \label{fig:interarrivalTimesAll}
\end{figure}

In this section, we show the performance of the random forest classifier considering only one feature, i.e., the interarrival time between consecutive packets.
For each device and service, we extracted and analyzed the inter-arrival times as reported in Figure~\ref{fig:interarrivalTimesAll}, which shows the interpolated inter-arrival times for the three devices and two services (except Google Nest Mini running one service). Although the samples have similar patterns, the interpolating polynomial functions have distinct patterns meaning that each service on each device can be potentially identified by just comparing the interarrival times. 

\vspace{-.3cm}

\vspace{-.3cm}

\begin{figure}[H]
\centering
  \begin{subfigure}[b]{0.495\textwidth}
    \includegraphics[width=\textwidth]{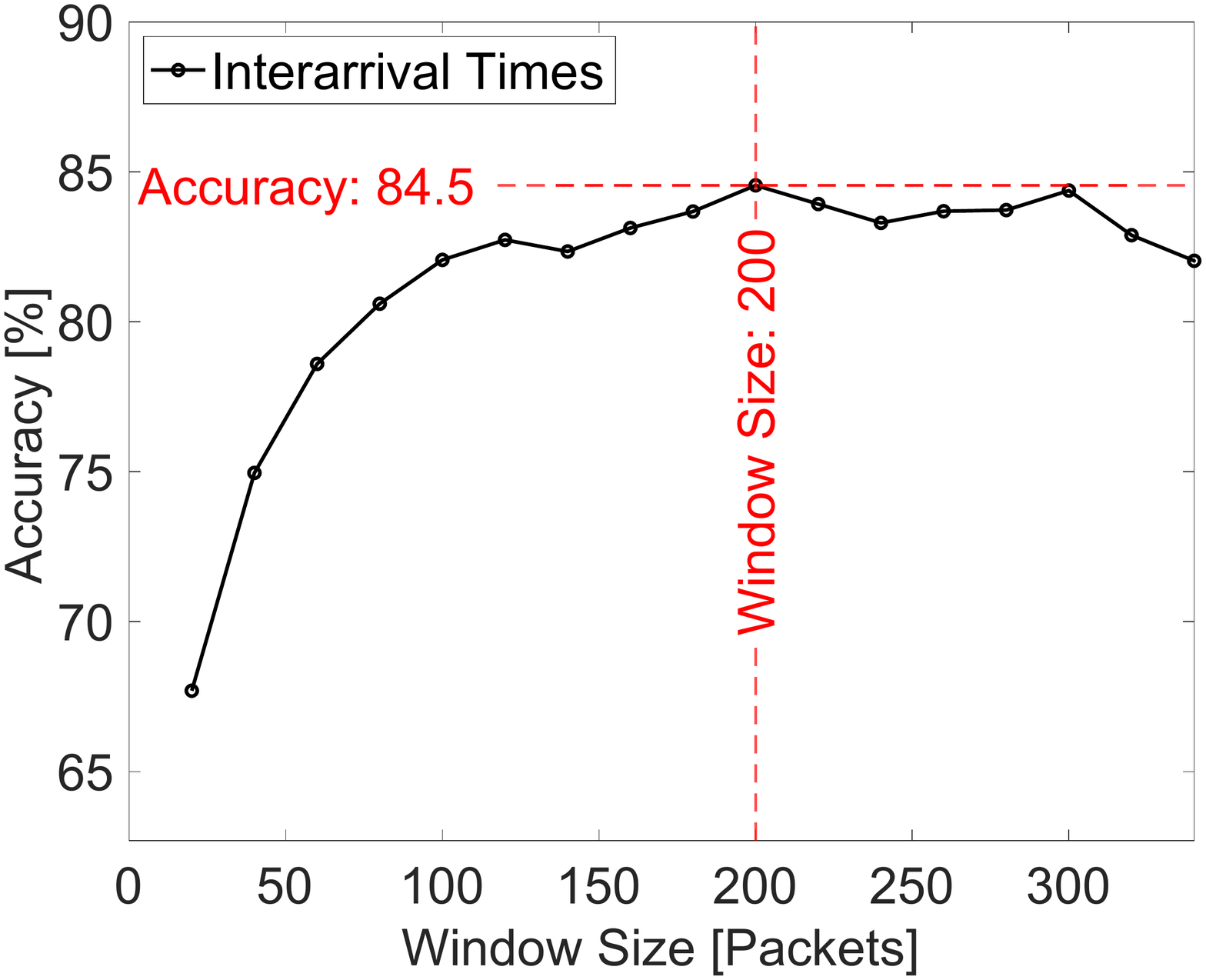}
    \caption{}
  \end{subfigure}
  \begin{subfigure}[b]{0.495\textwidth}
    \includegraphics[width=\textwidth]{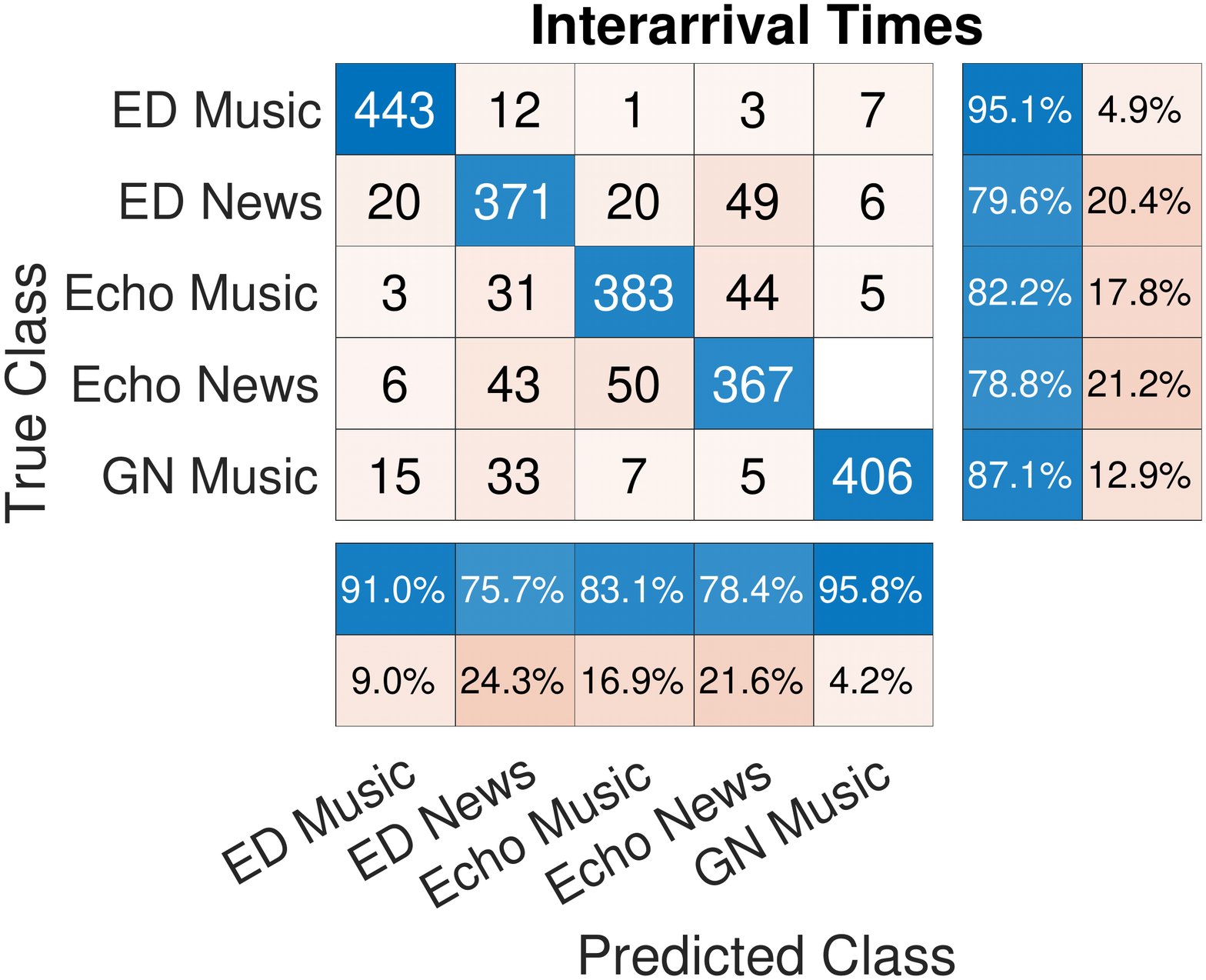}
    \caption{}
  \end{subfigure}
  \caption{Random Forest Classifier performance: (a) Accuracy as a function of Window Size. (b) Confusion Matrix associated with window size of 200 and 8 statistical features.}%
  \label{fig:wndSizeIAT}%
\end{figure}

\noindent{\bf Classification.} We consider a naive approach and perform the classification of the encrypted traffic by resorting only to the interarrival times achieving an accuracy of about 33.59\%. Subsequently, we considered the previously introduced statistical features computed from the inter-arrival times using the \emph{sliding window} technique. The window size has been set between 20 to 340 packets. Figure~\ref{fig:wndSizeIAT} (a) shows the accuracy of the Random Forest algorithm as a function of the window size. We observe that the trend of the accuracy saturates when it reaches the value of 200 being equivalent to about 80 seconds. We highlight that we choose the shortest window size (with the highest possible accuracy), in this case 200 packets, to guarantee the shortest detection delay.
Figure~\ref{fig:wndSizeIAT} (b) shows the confusion matrix for the previously considered five classes, i.e., Amazon Echo Dot Music (ED Music), Amazon Echo Dot News (ED News), Amazon Echo Music (Echo Music), Amazon Echo News (Echo News), Google Nest Music (GN Music), while the total number of samples is 2330. We considered the Random Forest algorithm and the 8 statistical features previously introduced, estimated over a window size of 200 samples that is sufficient and leads to an accuracy of 84.5\%.

\vspace{-.3cm} 

\vspace{-.7cm} 

\subsection{Packet Size Analysis}
\begin{figure}[H]
    \centering
    \includegraphics[width=0.7\textwidth]{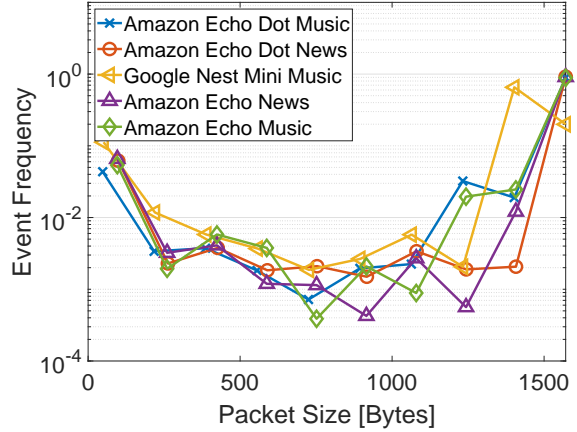}
    \caption{Probability distribution function associated with packet sizes.}
    \label{fig:packetSizes}
\end{figure}

Packets sizes can be easily extracted and used directly for analysis (to identify the existence of a unique pattern) and classification. Figure~\ref{fig:packetSizes} shows the packet size analysis as a function of the five classes considered in this paper. We observe that the frequency associated with the packet size is very different among the five considered categories, in particular when considering packet sizes larger than 800 bytes.

\vspace{0.25cm}

\begin{figure}[H]
\centering
  \begin{subfigure}[b]{0.475\textwidth}
    \includegraphics[width=1\textwidth]{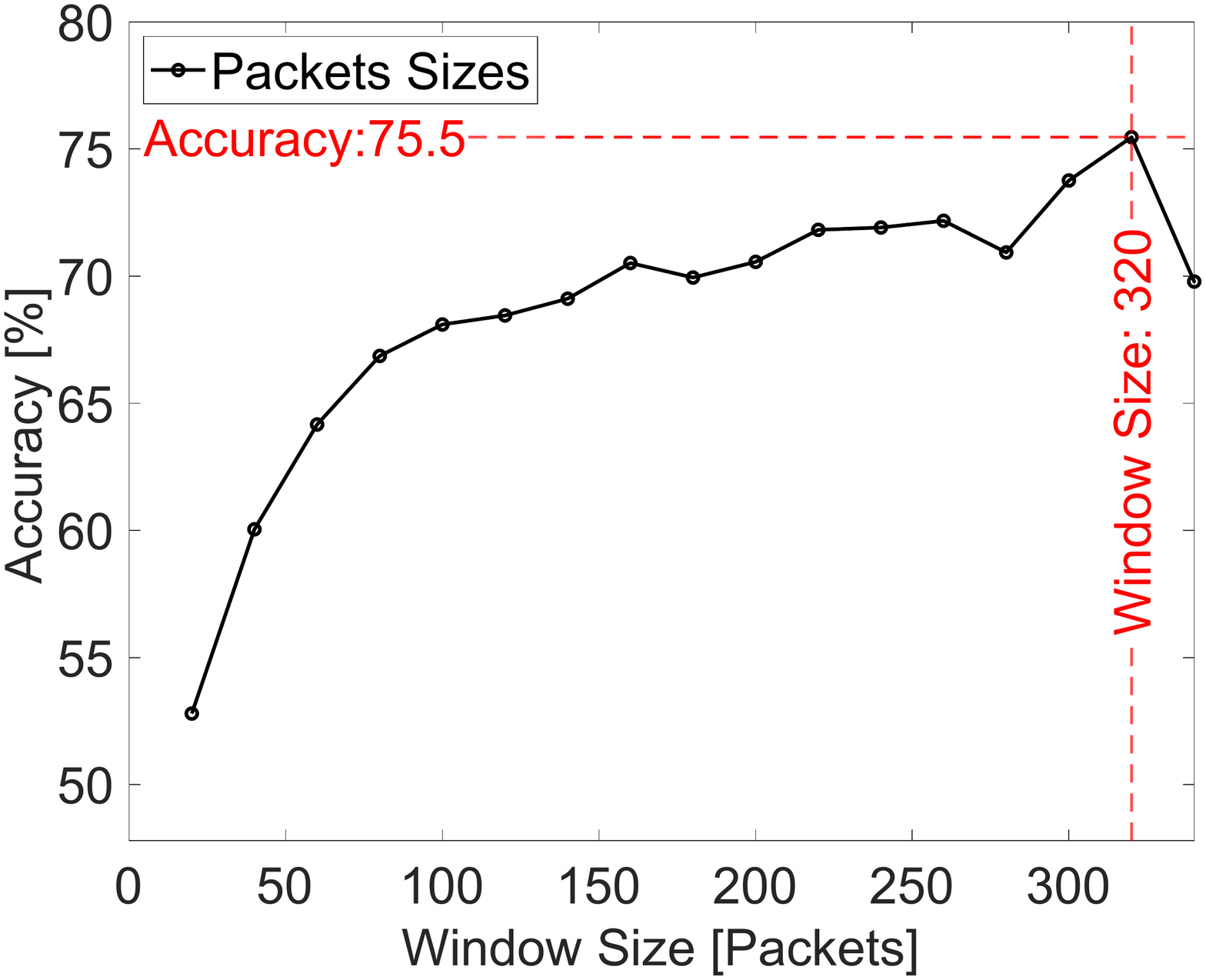}
    \caption{}
  \end{subfigure}
  \begin{subfigure}[b]{0.475\textwidth}
    \includegraphics[width=1\textwidth]{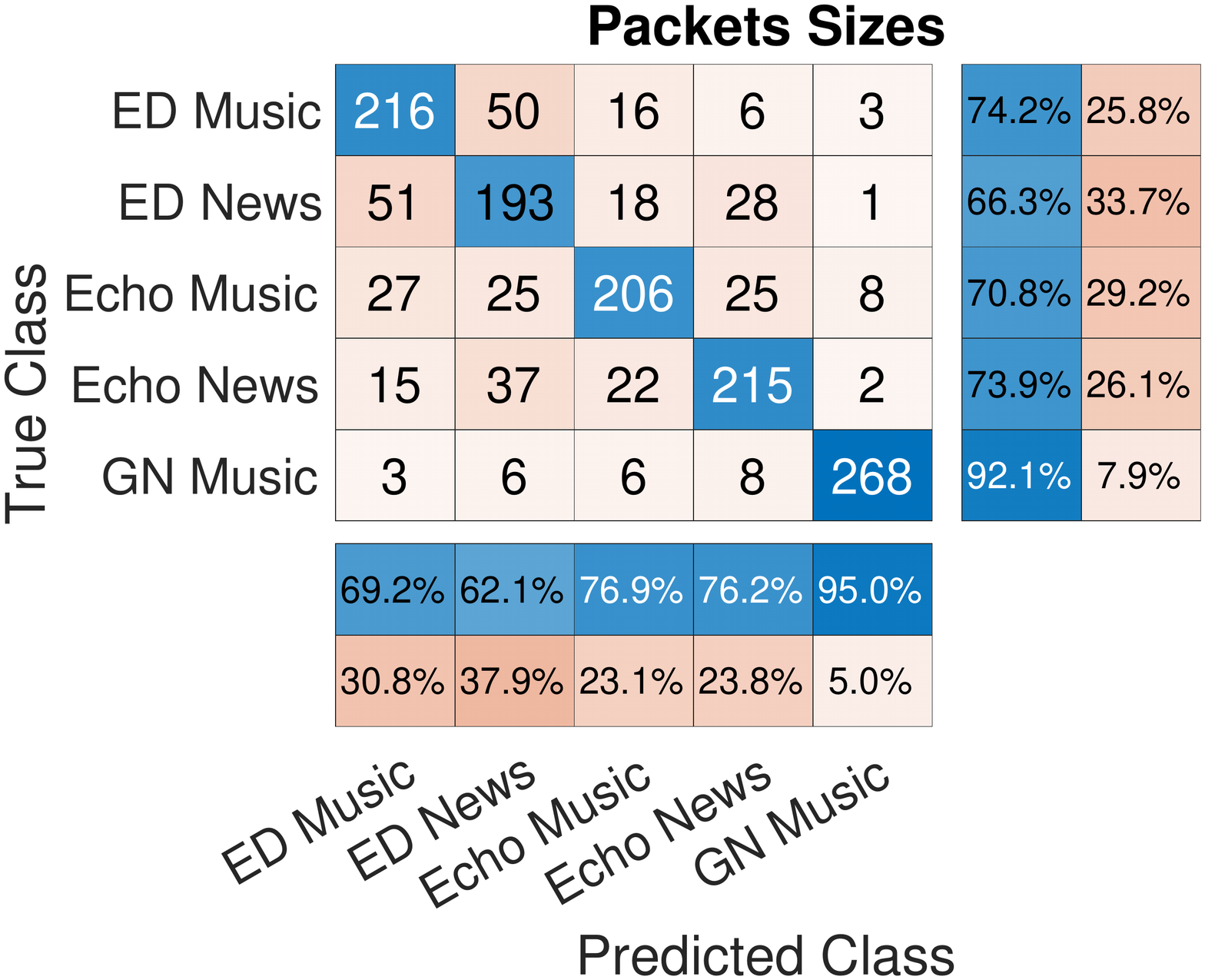}
    \caption{}
  \end{subfigure}
    \caption{Random Forest Classifier performance: (a) Accuracy as a function of Window Size. (b) Confusion Matrix associated with window size of 320 and 8 statistical features.}%
    \label{fig:wndSizePZ}%
\end{figure}

\noindent{\bf Classification.} As a baseline scenario, we consider the Random Forest algorithm and only the packet size feature. As for the previous case (interarrival times), such configuration gives us very poor performance, i.e., the accuracy achieved is 
0.33. Therefore, we consider 8 statistical features computed using the sliding window technique with a range between 20 and 340 subsequent packets. Figure~\ref{fig:wndSizePZ} (a) shows the accuracy of the Random Forest algorithm as a function of the packet size considering different window sizes. The highest accuracy achieved is approximately 0.755, using a window size of 320 packets being equivalent to about 130 seconds. Figure~\ref{fig:wndSizePZ} (b) shows the confusion matrix for the previously considered five classes, i.e., Amazon Echo Dot Music (ED Music), Amazon Echo Dot News (ED News), Amazon Echo Music (Echo Music), Amazon Echo News (Echo News), Google Nest Music (GN Music), where the total number of samples is 1455. We considered the Random Forest algorithm and the 8 statistical features previously introduced estimated over a window size of 320 samples.

\subsection{Packet Size and Interarrival Times}
\label{sec:packet_size_inttime}
\begin{figure}[H]
\centering
  \begin{subfigure}[b]{0.495\textwidth}
    \includegraphics[width=\textwidth]{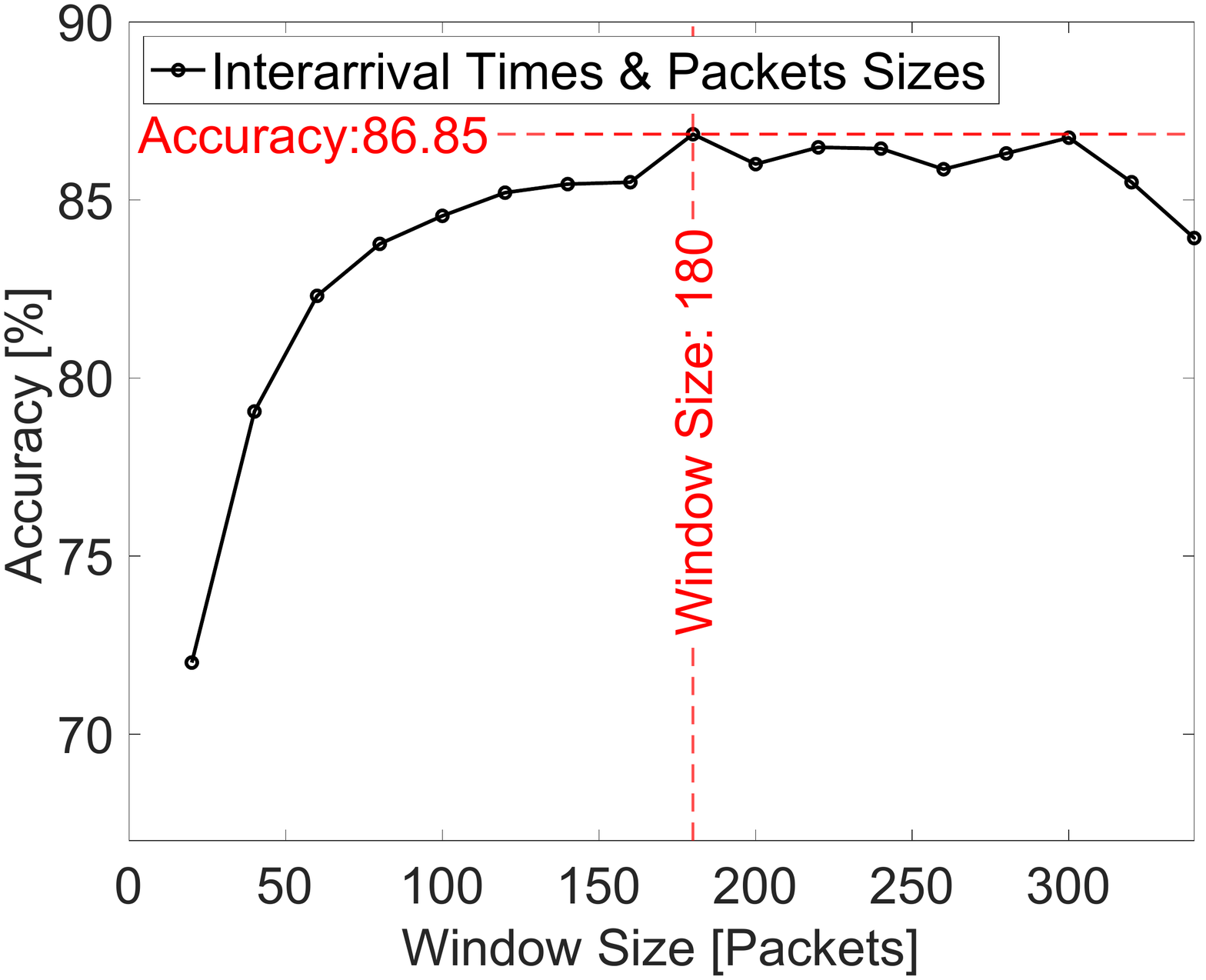}
    \caption{}
  \end{subfigure}
  \begin{subfigure}[b]{0.495\textwidth}
    \includegraphics[width=\textwidth]{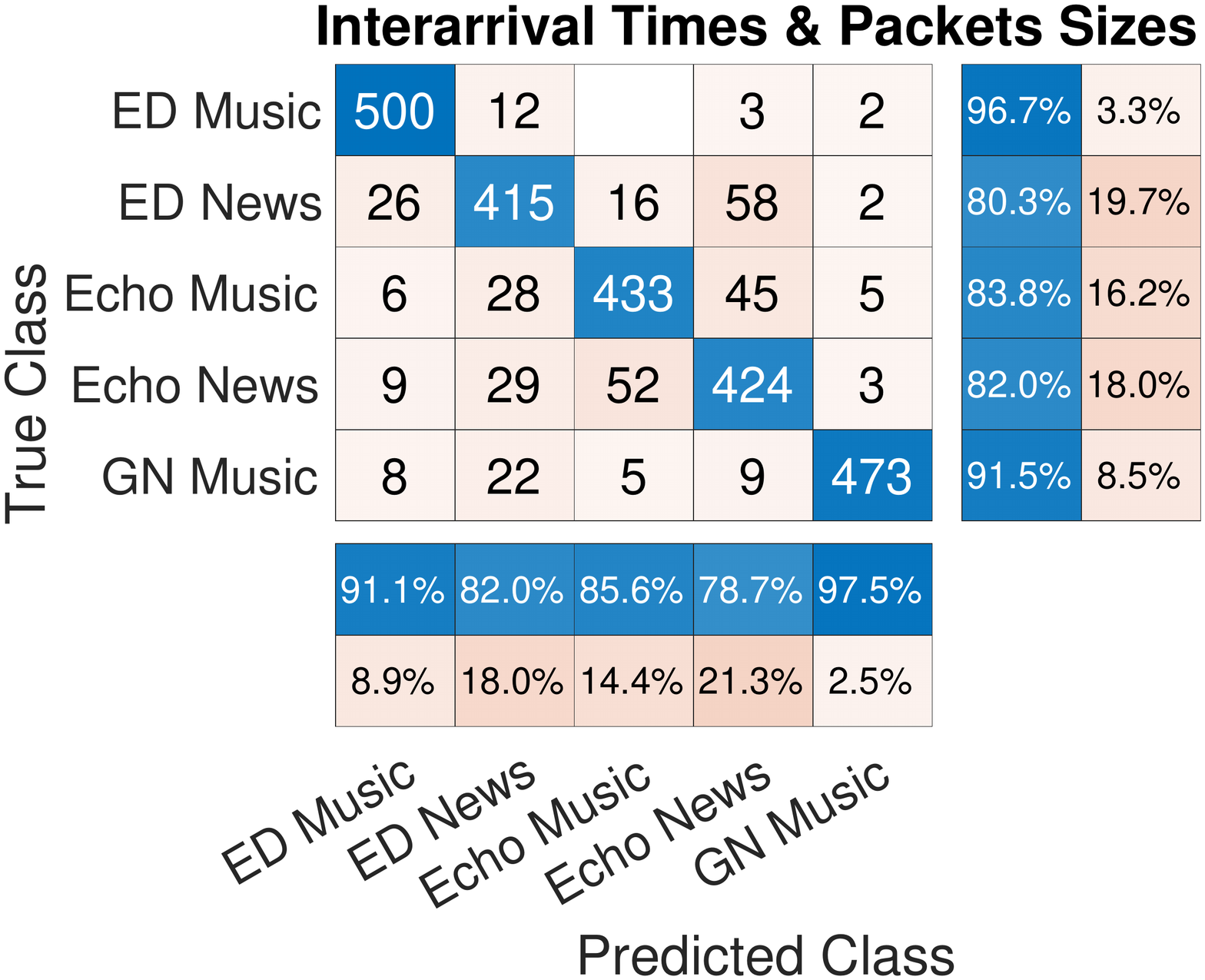}
    \caption{}
  \end{subfigure}
    \caption{Random Forest Classifier performance: (a) Accuracy as a function of Window Size. (b) Confusion Matrix associated with window size of 180 and 16 statistical features.}%
    \label{fig:iatpzwnd}%
\end{figure}

In this section, we consider the combination of packet size and interarrival times, by combining all the statistical features previously computed together, i.e., 8 for the packet size and 8 for the interarrival time for a total of 16 values. As for the previous cases, we consider the Random Forest algorithm and a sliding window spanning between 20 and 340. Figure~\ref{fig:iatpzwnd} (a) shows the accuracy as a function of the window size considering a total of 16 features. The accuracy saturates to the value of about 86\% without any major improvement when the window size is becoming greater than 180 being equal to 80 seconds. While Figure~\ref{fig:iatpzwnd} (b) shows the confusion matrix (with a total of 2585 samples) associated with the performance of the Random Forest algorithm when considering 16 statistical features computed from both packet size and interarrival time as previously described.

\vspace{-.8cm}

\section{Device and Service Detection in the Wild}
\label{sec:device_detection}
While in the previous sections we focus on devices and services identification, in this section, we show the performance of the Random Forest classifier to detect the presence of the considered smart devices and services running, from a crowded WiFi traffic.
The noise is collected from an environment where different number of users are connected to the same access point. The noise includes different types of packets, such DNS queries, ACK, ICMP, HTTP, HTTPs, QoS.
As previously considered, we start our analysis by providing a statistical description of the WiFi traffic we collected from a crowded place. Firstly, we faced a class imbalance problem; indeed, the WiFi trace collected from the wild is much larger than the one collected from the devices. In order to make the comparison fair, we divided the WiFi trace in chunks of the same size of the trace collected from each device associated with each service, once detecting the device type. In the case of having two devices of the same type running the exact same service within the WiFi channel, the classifier will match and classify both devices to the same class.
We applied the previously introduced methodology and we trained a model to identify each of the 5 classes, i.e., Amazon Echo Dot (Music), Amazon Echo Dot (News), Amazon Echo (Music), Amazon Echo (News), Google Nest Mini (Music), against the WiFi collected from the crowded place. Finally, we tested the model with independent traces taken from the previous 5+1 classes of traffic. We run the classifiers on each chunk and we averaged the final results. 
Table~\ref{tab:TracesVsNoise} shows the results associated with the detection of each of the device/service previously considered assuming a window size of 180 samples. We report performance in terms of True Positive Rate (TPR), False Positive Rate (FPR) and Accuracy. Our results show that performance is striking and our approach can easily detect a smart device and the associated service from a generic encrypted WiFi link.

\begin{table}[H]
\centering
\caption{Performance for Devices/Services detection}
\label{tab:TracesVsNoise}
    \begin{tabular}{|c|c|c|c|c|}
        \hline
        \textbf{Device} & \textbf{Service} & \textbf{TPR} & \textbf{FPR} $[\cdot 10^{-4}]$ & \textbf{Accuracy} \\ \hline
        \multirow{2}{*}{Amazon Echo Dot} & Music & 0.9988 &  3.4 & 0.9992 \\ \cline{2-5} 
        & News & 0.9996 &  7.9 & 0.9994 \\ \hline
        \multirow{2}{*}{Amazon Echo} & News & 0.9996 & 2.7 & 0.9996 \\ \cline{2-5}
        & Music & 0.9998 & 3.6 & 0.9997 \\ \hline
        Google Nest Mini & Music & 0.9995 & 9.2 & 0.9997 \\ \hline
    \end{tabular}
\end{table}

\noindent{\bf Comparison with other adversarial models.} Other attacks introduced in literature focus on using either raw or statistical features extracted from multiple information found in the exchanged packets such as (but not limited to) port number, packet direction, round trip time (RTT), inter-packet time (IPT), and other none-encrypted information. Our attack only leverages the two aforementioned features, interarrival times and packet size.
Table~\ref{tab:comparison} shows the comparison between our solution and the most representative in the literature in terms of type of device, technology, connectivity, attack strategy and capability to classify the application associated to the eavesdropped traffic. We remark that this contribution is the first to deal with encrypted traffic classification over a wireless link and considering only one flow direction, i.e., from the AP to the IoT device(s). Previous work~\cite{sivanathan2017characterizing} considered the same problem but taking into account a wired link making their scenarios (wired) not comparable with ours (wireless)---the wireless scenario being more challenging, indeed the adversary can be stealthier since he might hide himself behind obstructions, while the radio link is characterized by delays and jitters that make the classification process more difficult, but as we proved in this paper, achievable with overwhelming probability.
Moreover, to the best of our knowledge, other contributions do not differentiate between services and devices, while in our analysis we provide the likelihood to detect a specific device, and subsequently, to identify the provided service.

\begin{table}[H]
	\centering
	\caption{Comparison of our approach with other solutions.}
	\label{tab:comparison}
	\begin{tabular}{|c|>{\centering\arraybackslash}p{2.3cm}|>{\centering\arraybackslash}p{2cm}|>{\centering\arraybackslash}p{2.5cm}|>{\centering\arraybackslash}p{2cm}|>{\centering\arraybackslash}p{2cm}|>{\centering\arraybackslash}p{4cm}|>{\centering\arraybackslash}p{2cm}|}
		\hline
		\textbf{Ref.} & \textbf{WiFi Devices Only}  & \textbf{Remote Attack} & \textbf{Services Classification}  & \textbf{Single Direction Flow} \\ \hline
		\cite{msadek2019iot} &  \xmark & \cmark & \xmark & \xmark\\ \hline
		\cite{shahid2018iot} &  \xmark & \xmark & \xmark & \xmark \\ \hline   
		\cite{santos2018efficient} &  \xmark & \cmark & \xmark & \xmark \\ \hline
		\cite{sivanathan2017characterizing} & \xmark &  \cmark & \cmark & \xmark \\ \hline
		\cite{jackson2018amazon} & \cmark & \cmark & \cmark & \xmark \\ \hline
		Our Approach & \cmark & \cmark & \cmark & \cmark \\ \hline
	\end{tabular}
\end{table}

\section{Our Countermeasure: Eclipse}
\label{sec:countermeasure}
In the following, we propose an effective and efficient solution to the aforementioned attack. Our intuition is that encrypted network traffic classification can be made (arbitrary) less effective if the features characterizing the different traffic flows are reshaped, thus becoming similar. Indeed, the performance of the attack can be significantly mitigated if the adversary is not able to build a ground-truth related to the device/service traffic. We propose a simple but effective solution we name \emph{Eclipse}. Eclipse is implemented in the WiFi access point as a proxy between the  Internet traffic (wired) and the wireless connections between the WiFi access point and the IoT devices (recall Fig.~\ref{fig:attack_setup}). 
We adopt our previously introduced classification methodology by taking into account 16 total features (8 for the packet size and 8 for the interarrival time), considering a window size of 180 packets.

\vspace{0.25cm}

\begin{figure}[H]
\centering
  \begin{subfigure}[b]{0.55\textwidth}
    \includegraphics[width=\textwidth]{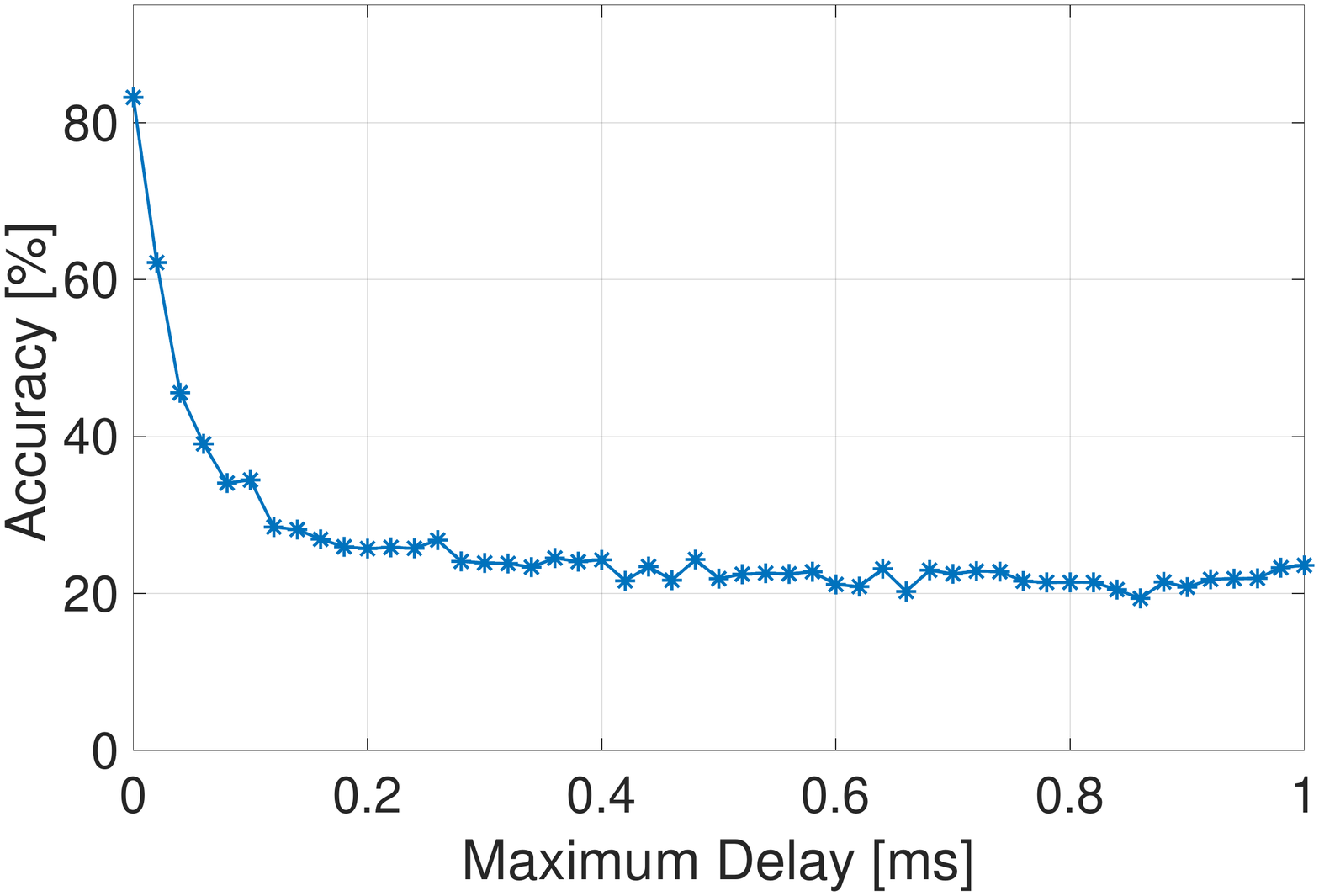}
    \caption{}
  \end{subfigure}
  \begin{subfigure}[b]{0.55\textwidth}
    \includegraphics[width=\textwidth]{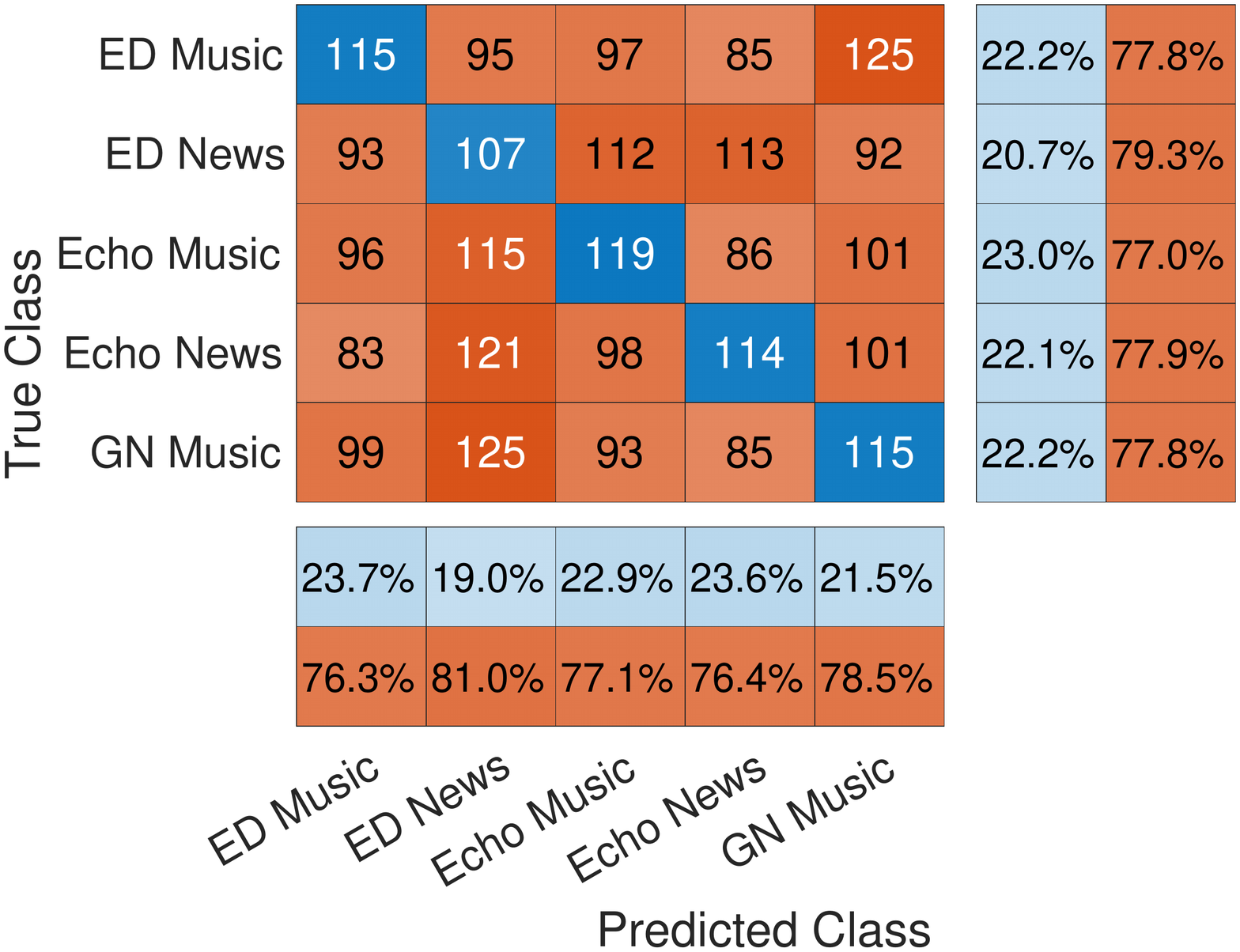}
    \caption{}
  \end{subfigure}
    \caption{(a) Classification accuracy as a function of the introduced mitigation method. (b) Confusion Matrix associated with the obscured data.}%
    \label{fig:AccuracyVsDelay}%
\end{figure}

\noindent{\bf Interarrival time and packet size reshaping.} We implement the traffic reshaping for both the packet size (as previously discussed) and for the interarrival times, i.e., by introducing random delays in the packet forwarding. Figure~\ref{fig:AccuracyVsDelay} shows the classification accuracy as a function of the introduced (maximum) delay. We observe how the overall accuracy drops from about 85\% to about 20\% by rescheduling the packets with a (maximum) delay of 0.5 ms.
We highlight that training a new model with the reshaped traffic is useless since the features are no more biased towards a specific class (traffic flow), thus making the training (on the reshaped traffic) completely ineffective.

\newpage

\section{Conclusion}
\label{sec:conclusions}
In this paper, we have presented a methodology to identify and classify popular smart devices and related streaming services by adopting standard machine learning algorithms. Our approach is based on extracting and detecting network patterns (features) such as packet size and interarrival times and the related statistics. We prove that our attack is feasible and can detect and identify smart devices with probabilities, about 0.99  and 0.86, respectively. We have proved that, contrary to the common belief, multi-layer encryption does not solve the problem of privacy, highlighting that smart devices require additional attention to guarantee the privacy of the end-user. Finally, we proposed a novel mitigation technique that prevents the identification of the traffic pattern by reshaping both the packet size and the inter-arrival times. Our findings prove that reshaping the packet size, e.g., setting all the packets to the same size, is completely ineffective to hide the presence of a specific flow, while introducing (small) random delays in the packet forwarding process makes the identification of the flow comparable to the random classification baseline. 

\section*{Acknowledgments}
This publication was made possible by NPRP grants NPRP12S-0125-190013 and NPRP12C-0814-190012 from the Qatar National Research Fund (a member of Qatar Foundation). The findings achieved herein are solely the responsibility of the authors.

\bibliographystyle{IEEEtran}
\bibliography{thedarksideofiot_spiot2020}

\balance
\end{document}